\documentstyle[12pt]{article}
\date{}
\begin{document}
\rightline{\bf CU-TP-950}

\begin{center}
{\Large\bf The Boltzmann equation for gluons at\\
 \indent early times after a heavy ion collision}

\vskip 50pt
A.H. Mueller\footnote{This work is supported in
Part by the Department of Energy.}\\
Physics Department, Columbia University\\
New York,  N.Y. 10027
\end{center}

\noindent{\bf Abstract}
\vskip 20pt
A Boltzmann equation is given for the early stages of evolution of the
gluon system produced in a head-on heavy ion collision.  The collision term is
taken from gluon-gluon scattering in the one-gluon approximation. 
$<p_\perp>$ and $<p_z^2>$ are evaluated as a function of time using
initial conditions taken from the McLerran-Venugopalan model.

\section{Introduction}

In an earlier paper\cite{Mue} we studied the early stages of the gluon
system produced in a heavy ion collision. In that paper we took the
initial conditions of the gluon system from the
McLerran-Venugopalan model assuming that all gluons at or below the
saturation momentum in the light-cone wavefunction of the
McLerran-Venugopalan model are freed in the heavy ion collision while
all gluons beyond the saturation momentum are not
freed\cite{Ler,Nuc,Jal,Kov,Yu}.  We then wrote an equation for the rate
of change of the $typical$ transverse momentum of a gluon in the
system as a function of time.  This equation takes into account, in an
average way, the transfer of transverse momentum into longitudinal
momentum due to elastic gluon-gluon scattering in the one gluon exchange
approximation.  We were able to follow the decrease of the typical
transverse momentum of a gluon with time up to times on the order of
${1\over Q_s} exp[const/{\sqrt{\alpha}}\ ]$  at which time the typical
transverse and longitudinal momenta become equal, and our procedure broke
down.  We argued that this was likely the time at which equilibrium was
setting in and our simple approach was not capable of following the gluon
system into equilibrium.

In this note we reformulate our previous discussion in a much more
general way in terms of the Boltzman equation\cite{Lif} with the
collision term taken from elastic gluon-gluon scattering in the one gluon
exchange approximation.  In the logarithmic approximation for integrating
over the angle of scattering, the same approximation used in Ref.1, the
Boltzmann equation reduces to a Fokker-Planck equation\cite{Lif}, given
in (15) and (16), for the number of gluons, f, per unit of phase space.  The
infrared cutoff on very small angle scattering is taken from Ref.1
although the exact form of this cutoff is not crucial for the existence
of an equation of the type given in (15).

There are two explicit parameters appearing in (15).  The first of these
parameters, $\eta,$ which is time-independent is given by the initial
conditions for the evolution.  In our explicit evaluations we take
$\eta$ from the McLerran-Venugopalan model in terms of the saturation
momentum $Q_s.$  The second parameter, $\eta_{-1},$ given in (13) should
be determined self-consistently from the solution to (15).  In our
explicit evaluations we take $\eta_{-1}$ to be given by the
McLerran-Venugopalan model at very early times (28), while the time
dependence of $\eta_{-1}$ can be estimated using (32).  We have also
assumed $t_o,$ in (10), to be time-independent.  In Ref.1, we found the
time-dependence of $t_o$ to have a negligible effect at early times and
we expect the same to be true here.  However, before attempting to solve
(15) more completely, say by numerical means, at non early times one must
give a specification of $t_o,$ in (10), analogous to what was done in
Ref.1, but now in the context of the Boltzmann equation.

\section{The Boltzmann equation}

A derivation of the Boltzmann equation can be found in Ref.7, and can be
written as

\begin{equation}
{\partial f\over \partial t} + \vec{v} \cdot \bigtriangledown_{\vec{x}} f= C
\end{equation}

\noindent where\   C\ is the collision term.  $f$  is the number of gluons per
unit of phase space after the valence quarks have separated in a high energy,
head-on, heavy ion collision.  Thus $f=f(\vec{p},\vec{x}, t)$ and

\begin{equation}
N(\vec{x},t) = \int d^3 \vec{p}f
\end{equation}

\noindent is the gluon density per unit volume.  $f$  is a scalar function. 
We shall assume, for simplicity, that $f$ only depends on z, the coordinate
corresponding to the axis along which the heavy ions collide, but not on the
transverse coordinates $\underline{x}.$  Further we suppose that the physics
of the heavy ion collision is boost invariant, at least for final state
particles not having too large a rapidity in the center of mass system.  If we
parameterize the momentum and position of a final state gluon by

\begin{equation}
p_\mu=p_{_\perp}(cosh\  y, \b{\^v}, sinh\  y) = (p_0,\underline{p}, p_z)
\end{equation}
\begin{equation}
x_\mu = \tau (cosh\  \eta, \b{0}, sinh\  \eta) = (t, \b{x}, z)
\end{equation} 

\noindent then $f = f(p_\perp, \tau, y-\eta)$ expresses the boost invariance
of the distribution of particles in the final state.  In terms of $\tau, y,\eta$

\begin{equation}
({\partial\over \partial t} + \vec{v}\cdot \bigtriangledown_{\vec{x}}) f = ({\partial\over \partial
t}+ v_z {\partial\over \partial z})f = {cosh(y-\eta)\over cosh y}({\partial\over
\partial t} - {tanh(y-\eta)\over
\tau} \ {\partial\over \partial y})f
\end{equation}

\noindent or, taking $\eta = z = 0$

\begin{equation}
({\partial\over \partial t} + v_z{\partial\over \partial z}) f =
({\partial\over \partial t} - {p_z\over t} {\partial\over \partial p_z}) f(t,
p_\perp,p_z)
\end{equation}

\noindent where we have expressed, at $z=0,\ f$ as a function of $t, p_\perp$
and $p_z,$ variables convenient for the discussion to follow. 

Now  turn to the collision term in (1).  It is straightforward to show that

\begin{equation}
C = \partial_i\int d^3p_2[(\partial_j-\partial_{2j}) ff_2] B_{ij}
\end{equation}

\noindent where we use the notation $\partial_i={\partial\over \partial p_i}, \partial_{2i} =
{\partial\over \partial p_{2i}}, f =f(t,p_\perp,p_z), f_2=f(t, p_{2\perp},p_{2z}).\  B_{ij}$ is
given by

\begin{equation}
B_{ij} = 2\pi\alpha^2({N_c^2\over N_c^2-1}) L[(1-\vec{v}\cdot \vec{v}_2) \delta_{ij} + v_iv_{2j}+
v_{2i} v_j)]
\end{equation}

\noindent with $\vec{v} = {\vec{p}\over \vert\vec{p}\vert}$ and $\vec{v}_2 =
{\vec{p}_2\over \vert \vec{p}_2\vert}.$  Except for the factor $({N_c^2\over
N_c^2-1})$ the $B_{ij}$ in (8) is identical to that given in Ref.7 for QED.  The factor \  L\
represents a (perturbatively divergent) integral over the (small) angle of scattering which we cut
off as in Ref.1 at a minimum scattering angle $\theta_m$ given by

\begin{equation}
\theta_m^3 = {t_0\over t}
\end{equation}

\noindent with $t_0$ slowly varying with time for $t$ not too large.  We shall treat $t_0$ as a
constant in this paper.  Thus,

\begin{equation}
L = \int_{\theta_m}^1{d\theta\over \theta} = {1\over 3} \ell n\  t/t_0 \equiv {1\over 3}  \xi.
\end{equation}

Using (8) and (10) in (7) along with (1) and (6) leads to 

\begin{equation}
C = \lambda N \xi \bigtriangledown_{\vec{p}}^2 f((t,\vec{p}) + 2 \lambda N_{-1}\xi
\bigtriangledown_{\vec{p}}\cdot(\vec{v}f(t,\vec{p}))
\end{equation}

\noindent where

\begin{equation}
N(t) \int d^3p f(t,\vec{p})
\end{equation}

\noindent is the gluon number density while

\begin{equation}
N_\gamma(t) = \int d^3p f(t,\vec{p}) \vert \vec{p}\vert^\gamma
\end{equation}
\noindent and where
\begin{equation}
\lambda = {2\pi\alpha^2\over 3} ({N_c^2\over N_c^2-1}).
\end{equation}

\noindent Defining $tN=\eta, tN_{-1} = \eta_{-1}$ and using (6) and (11) in (1) finally leads to an
equation of the Fokker-Planck type 
\begin{equation}
({\partial\over \partial\xi} - {\partial\over \partial p_z} p_z)
\tilde{f}=\lambda\eta\xi \bigtriangledown_{\vec{p}}^2\tilde{f} + 2\lambda\eta_{-1}\xi
\bigtriangledown_{\vec{p}}(\vec{v}\tilde{f})
\end{equation}

\noindent with

\begin{equation}
t f(t,\vec{p}) = \tilde{f}(t,\vec{p}).
\end{equation}
\section{Properties of the solution of the Boltzmann equation}

It appears difficult to solve (15) even with simple initial conditions.  Nevertheless, one can
discern general characteristics of the solutions, for early times, even without being able to
completely solve the equation.  Taking the integral with respect to $d^3p$ on both sides of (15)
and recalling that $\int \tilde{f} d^3p = \eta$ one finds

\begin{equation}
{\partial\over \partial\xi} \eta = 0
\end{equation}

\noindent which means that the number density of gluons $N(t)$ decreases like $1/t.$

Multiplying both sides of (15) by $p_\perp^2$ and integrating over all $\vec{p}$ leads to

\begin{equation}
{\partial\over \partial\xi} <p_\perp^2> = 4\lambda\eta\xi(1-{\eta_{-1}\eta_{+1}\over \eta^2}) +
{4\lambda\eta_{-1}\xi\over \eta} \int d^3p \tilde{f}
{p_z^2\over p}
\end{equation}

\noindent while multiplication by $p_z^2$ and integration yields

\begin{equation} 
{\partial\over \partial\xi} <p_z^2> + 2 <p_z^2> = 2\lambda\eta\xi - {4\lambda\eta_{-1}\xi\over \eta}
\int d^3p \tilde{f} {p_z^2\over p}.
\end{equation}

\noindent In (18) and (19) we have used

\begin{equation}
\int d^3p \tilde{f} p_\perp^2 \equiv <p_\perp^2> \eta
\end{equation}

\noindent and

\begin{equation}
\int d^3p \tilde{f} p_z^2 \equiv <p_z^2> \eta.
\end{equation}

\noindent In case $\tilde{f}$ does not decrease faster than $p_\perp^{-4},$ which is the behavior
in perturbative QCD, it may be useful to replace (18) by

\begin{equation}
{\partial\over \partial\xi} <p_\perp> = \lambda\xi(\eta_{-1}^\perp -
{2\eta_{-1}\eta(p_{\perp}/p)\over \eta})
\end{equation}

\noindent where

\begin{equation}
\eta_{-1}^\perp=\int d^3p \tilde{f} {1\over p_\perp}
\end{equation}

\noindent and

\begin{equation}
\eta(p_\perp/p) = \int d^3p \tilde{f} {p_\perp\over p}.
\end{equation}

\noindent At early times the values of $p_z$ are very small at $z=0$ so that we may
set $\eta_{-1}^\perp\approx \eta_{-1}$ and $\eta(p_\perp/p) \approx \eta$ so that (22) becomes

\begin{equation}
{\partial\over \partial\xi} <p_\perp> = 
- \lambda\xi\eta_{-1}.
\end{equation}

\noindent In the McLerran-Venugopalan model used in Ref.1

\begin{equation}
\tilde{f}_s(t,\vec{p}) = \delta(p_z) \Theta(Q_s^2-p_\perp^2) c{N_c^2-1\over 4\pi^3\alpha N_c}
\end{equation}

\noindent immediately after the ion-ion collision leading to

\begin{equation}
\eta = c{N_c^2-1\over 4\pi^2\alpha N_c} Q_s^2.
\end{equation}

\noindent While at early times

\begin{equation}
\eta_{-1} = c{N_c^2-1\over 2\pi^2\alpha N_c} Q_s
\end{equation}

\noindent so that (25) becomes

\begin{equation}
{\partial\over \partial\xi} <p_\perp> = - c{\alpha N_c\over 3\pi} Q_s\xi
\end{equation}

\noindent $Q_s$ is the saturation momentum.  Defining

\begin{displaymath}
\bar{Q} = {1\over \eta} \int d^2p \ p_\perp f_s = {2\over 3} Q_s
\end{displaymath}

\noindent one finds

\begin{equation}
<p_\perp/\bar{Q}> = 1 - c{\alpha N_c\over 4\pi} \xi^2
\end{equation}

\noindent a behavior having an identical form, for $\xi$ not too large, to that found in Ref.1.

Now turning to (19) it is clear that at early times the first term on the left-hand side of that
equation and the second term on the right-hand side are small compared to the remaining two terms at
early times so

\begin{equation}
<(p_z/\bar{Q})^2> = c{3\alpha N_c\over 4\pi} \xi.
\end{equation}

\noindent Over what range of $\xi$ can (30) and (31) be used?  Since we are working in a
logarithmic approximation\cite{Lif} we clearly need $\xi >> 1$ in order to take the collision term
to be given by (11).  In the McLerran-Venugopalan model $\xi \approx \ell n\  1/\alpha$ even when
$t$ is as small as $1/Q_s$ so the assumption of large $\xi$ should be (parametrically) good for
all times after the gluons are freed in the heavy ion collision.  In obtaining (25) from (22) we
have used $\eta_{-1}^\perp \approx \eta_{-1}$ and $\eta(p_\perp/p)\approx \eta.$  These assumptions
are good so long as typical values of $p_z$ are much less than typical values of $p_\perp.$  We
shall come back in a moment to determine when $\vert p_z\vert \approx p_\perp.$  In solving (25) we
have taken $\eta_{-1}$ to be constant, and given by (28) in the McLerran-Venugopalan model.  We
can estimate the corrections coming from the time variation of $\eta_{-1}$ by  taking

\begin{equation}
\eta_{-1}(t) = \eta_{-1} {\bar{Q}\over <p_\perp>}.
\end{equation}

\noindent If we use (32) in (25) the result

\begin{equation}
<p_\perp/\bar{Q}>^2 = 1 -c{\alpha N_c\over 2\pi} \xi^2
\end{equation}

\noindent emerges.  Eqs.33 and 30 are identical when $\alpha \xi^2 << 1,$ however, we can expect
(33) to be reasonable even when $\alpha \xi^2$ is not small.  Thus, we suggest that (31) and (33)
should be good estimates for the typical transverse and longitudinal momenta so long as $<p_z^2>$
remains less than $<p_\perp>^2.$  (The precise factor in front of the ${\alpha N_c\over 2\pi}
\xi^2$ in (33) is not expected to be reliable, however, because (32) should be expected to be a
reasonable estimate though not a precise evaluation.)

From (31) and (33) we see that $<p_z^2>$ and $<p_\perp>^2$ become comparable when $\xi= \xi_1$ with
$\xi_1$ given by

\begin{equation}
\xi_1={\sqrt{{2\pi\over c\alpha N_c}}} + 0(\alpha^0).
\end{equation}

\noindent When $\xi \geq \xi_1$ we can no longer expect (31) and (33) to be indicative of the
evolution of the gluon system after the heavy ion collision.  Eqs.(18), (19) and (22) should
still be correct as they follow from (15), but when $\xi \geq \xi_1$ we are no longer able to
estimate the terms appearing on the right-hand sides of (18), (19) and (22).  While we are unable
to follow the system, analytically, beyond $\xi = \xi_1$ we presume that typical values of
$p_\perp$ and $p_z$ remains equal, at $z=0,$ and that the gluonic system continues to evolve toward
equilibrium.

\end{document}